\documentclass[aps,prl,10pt,twocolumn,floatfix,showpacs,superscriptaddress,longbibliography]{revtex4-1}

\usepackage{graphicx,amsmath,amsfonts,bm}
\usepackage[colorlinks=true, citecolor=blue, linkcolor=blue, urlcolor=blue]{hyperref}
\usepackage[all]{hypcap}

\begin{document}

\title{The chiral anomaly in real space}
\author{C. Fleckenstein}
\email{christoph.fleckenstein@physik.uni-wuerzburg.de}
\affiliation{Institute of Theoretical Physics and Astrophysics, University of W\"urzburg, 97074 W\"urzburg, Germany}
\author{N. Traverso Ziani}
\email{niccolo.traverso@physik.uni-wuerzburg.de}
\affiliation{Institute of Theoretical Physics and Astrophysics, University of W\"urzburg, 97074 W\"urzburg, Germany}
\author{B. Trauzettel}
\affiliation{Institute of Theoretical Physics and Astrophysics, University of W\"urzburg, 97074 W\"urzburg, Germany}
\begin{abstract}
The chiral anomaly is based on a non-conserved chiral charge and can happen in Dirac fermion systems under the influence of external electromagnetic fields. In this case, the spectral flow leads to a transfer of right- to left-moving excitations or vice versa. The corresponding transfer of chiral particles happens in momentum space. We here describe an intriguing way to introduce the chiral anomaly into real space. Our system consists of two quantum dots that are formed at the helical edge of a quantum spin Hall insulator on the basis of three magnetic impurities. Such a setup gives rise to fractional charges which we show to be sharp quantum numbers for large barrier strength. Interestingly, it is possible to map the system onto a quantum spin Hall ring in the presence of a flux pierced through the ring where the relative angle between the magnetization directions of the impurities takes the role of the flux. The chiral anomaly in this system is then directly related to the excess occupation of particles in the two quantum dots. This analogy allows us to predict an observable consequence of the chiral anomaly in real space.
\end{abstract}

\maketitle
The physics of relativistic fermions, described by the Dirac equation, inspired generations of physicists and gave rise to many astonishing discoveries, such as a bound soliton state of fractional charge 1/2 in the presence of a kinked background mass term, first predicted by Jackiw and Rebbi \cite{RJack}. It was subsequently shown by Kivelson and Schrieffer that this charge 1/2 can be understood as a sharp quantum observable \cite{SKivel}. Interestingly, Goldstone and Wilzcek extended the corresponding theory to a bound state of any fractional charge considering complex solitons \cite{JGold}. The first connection to condensed matter physics was introduced by Su, Schrieffer, and Heeger considering domain walls in polyacetylen chains \cite{AJHeeger}. With the discovery of topological insulators \cite{Hasan2010,Qi2011}, another broad connection from condensed matter physics to particle physics was established. As part of that, Qi, Hughes, and Zhang \cite{Zhang} demonstrated the analogy of the Jackiw-Rebbi model to a system of two magnetic impurities aligned along the helical edge of quantum spin Hall systems. These peculiar one-dimensional (1D) edge states of 2D topological insulators have recently been theoretically predicted \cite{Kane2005,Bernevig2006} and soon after experimentally observed \cite{Konig2007}.

\begin{figure}
\includegraphics[scale=0.32]{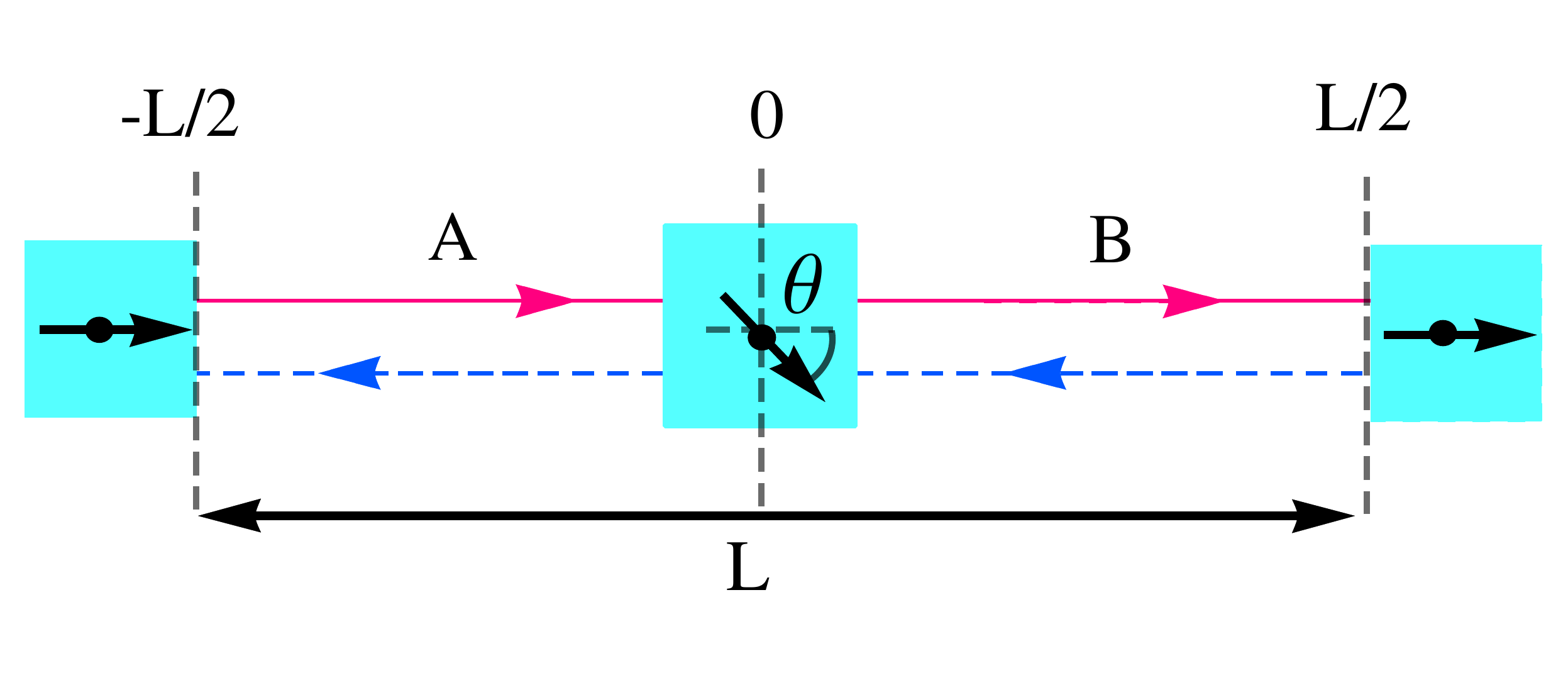}
\caption{(Color online) Schematic representation of the system: a helical edge containing three magnetic impurities, separating the system into regions A and B.}
\label{Fig:1}
\end{figure}

 Yet another prominent characteristic of relativistic fermions is the chiral anomaly, referring to a non-conservation of the chiral current, first studied by Adler \cite{SLAdler}, Bell, and Jackiw \cite{SBell} to explain the observed decay of a pion into two photons. With the theoretical prediction \cite{Volovik2003,Wan2011,Xu2011,Burkov2011a,Burkov2011b,ChaoXing} and experimental discovery of Weyl/Dirac semimetals \cite{Liu2014,Xu2015a,Xu2015b,JXiong}, the chiral anomaly has become prominent in condensed matter physics. In fact, it is considered to be one of the key features for an unambiguous  detection of Weyl points in Weyl semimetals. Since the non-conservation of right- and left-movers is intimately connected to momentum space, it is hard to get a direct experimental hold on this phenomenon. In fact, most theoretical predictions refer to indirect evidence, for instance, a pronounced negative magneto resistance. Thus, it is evidently desirable to predict more direct evidence of the chiral anomaly. Remarkably, we present a way to visualize this effect in real space, on the basis of an extended Jackiw-Rebbi model. Our model consists of three magnetic impurities along the helical edge, {\it{cf.}} Fig. \ref{Fig:1}. The two outer barriers are assumed to be aligned while the orientation of the inner one is rotated by an angle $\theta$ with respect to the outer ones. This choice of directions of the magnetization of the three impurities implements overlapping solitons and anti-solitons in our system. In the limit of large barrier strength, two exciting things can happen: (i) We identify sharp fractional charges in the two areas (i.e. quantum dots) between the impurities that can take any value. (ii) We predict that a real space analogy of the chiral anomaly can be observed if the orientation of the magnetization $\theta(t)$ of the center barrier is rotated as a function of time $t$.

\begin{figure}
\centering
\includegraphics[scale=0.38]{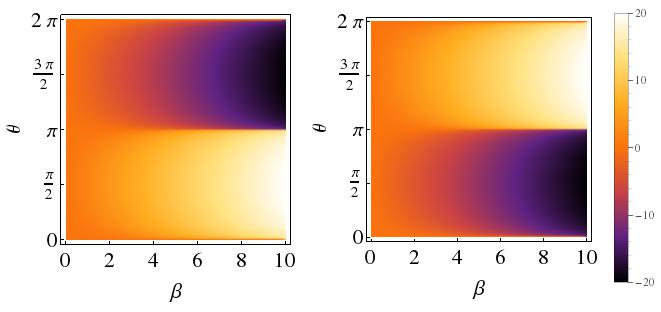}
\caption{(Color online) Plot of the relative occupations $\log(P^r_-(\beta,\theta))$ (left) and $\log(P^r_+(\beta,\theta))$ (right) as a function of the angle $\theta$ and the  magnetic strength $\beta=m/v_F$ of the central barrier.}
\label{Fig:DensityPlot1}
\end{figure}

The Hamiltonian $H$ of our system reads
\begin{equation}
\label{Eq:Hamiltonian}
H=\int_{-L/2}^{L/2}\hat{\Psi}^{\dagger}(x)\mathcal{H}(x)\hat{\Psi}(x)
\end{equation}
with the fermionic field operator $\hat{\Psi}(x)=(\hat{\psi}_{\uparrow}(x),\hat{\psi}_{\downarrow}(x))^T$, where the operators $\hat{\psi}_{\uparrow/\downarrow}$ annihilate $\uparrow/\downarrow$ particles. The Hamiltonian density can then be written as
\begin{equation}
\label{Eq:H01}
\begin{split}
\mathcal{H}(x)=-iv_F\sigma_z\partial_x-M\sigma_x\delta(L/2+x)\\
-M\sigma_x\delta(L/2-x)-m\left(\sigma_x\cos(\theta)-\sigma_y\sin(\theta)\right)\delta(x),
\end{split}
\end{equation}
where $\sigma_i$ ($i\in\lbrace x, y, z\rbrace$) are Pauli matrices acting on spin-space, $v_F$ is the Fermi velocity, and $M$, $m$ are parameters determining the strength of the magnetic impurities.
Following the formal ideas of Refs. \cite{1, 2}, we solve the time independent Schr\"odinger equation $\mathcal{H}(x)\Psi(x)=E\Psi(x)$
in the limit $M\rightarrow\infty$ and arbitrary $m$ and, thus, obtain the boundary conditions for the eigenfunctions $\Psi(x)$ at the points $x=0,\pm L/2$. The boundary conditions, explicitly written in App. A, motivate the following ansatz for the eigenfunctions
\begin{eqnarray}
\label{Eq:Ansatz}
\Psi_{n,\alpha}(x)=\Psi_{n,\alpha,A}(x)\theta(-x)\theta(x+L/2)\nonumber\\
+\Psi_{n,\alpha,B}(x)\theta(x)\theta(-x-L/2)
\end{eqnarray}
with
\begin{eqnarray}
\label{Eq:Ansatz1}
\Psi_{n,\alpha,\lambda}(x)=\begin{pmatrix}
a_{\lambda,\alpha}e^{iqx}\\
b_{\lambda,\alpha}e^{-iqx}\\
\end{pmatrix},
\end{eqnarray}
where $\lambda=(A,B)$. Evidently, $\Psi_{n,\alpha,A}(x)$ is located in region $A$ and $\Psi_{n,\alpha,B}(x)$ in region $B$. Using the boundary conditions and the wave functions of Eq. (\ref{Eq:Ansatz}), we obtain the quantization condition for the wave vector $q_{n,\alpha}$
\begin{equation}
\label{Eq:Quantization3Impurities}
q_{n,\alpha}=\phi_{\alpha}(\beta,\theta)+\frac{2\pi n}{L},~~~n\in\mathbb{Z},
\end{equation}
where $\phi_{\alpha}(\beta,\theta)=\alpha\frac{1}{L}\arccos(\tanh(\beta)\cos(\theta+\pi))$ with $\beta=m/v_F$ \cite{beta} and $\alpha=\pm$.
Putting $\hbar=1$, the corresponding eigenenergies are given by $E_{n,\alpha}=v_Fq_{n,\alpha}$.
The parameters $a_{\lambda,\alpha}$ and $b_{\lambda,\alpha}$ introduced in Eq. (\ref{Eq:Ansatz1}) are defined in App. A. To get information about the spatial distribution of the states $\Psi_{n,\alpha}(x)$ as a function of the angle $\theta$ and the magnetic strength $\beta$, we investigate the probability of finding a particle in the state $\Psi_{n,\alpha}(x)$ in region A and B, respectively.
For illustration, in Fig. \ref{Fig:DensityPlot1}, we plot the logarithm of the relative occupations
\begin{eqnarray}
P^r_{\alpha}(\beta, \theta)=\frac{P_{A,\alpha}(\beta,\theta)}{P_{B,\alpha}(\beta,\theta)},\nonumber
\end{eqnarray}
where we define
\begin{eqnarray}
P_{\lambda,\alpha}(\beta,\theta)=\lambda\int_{0}^{\lambda L/2}dx\Psi^{\dagger}_{n,\alpha}(x)\Psi_{n,\alpha}(x)\nonumber
\end{eqnarray}
with $\lambda=(B,A)=(+,-)$.
Notably, for small impurity strength $\beta$, arbitrary angle $\theta$, and also arbitrarily large values of $\beta$ and $\theta=n\pi$ ($n\in\mathcal{Z}$), the state $\Psi_{n,\alpha}$ is equally extended in regions A and B, with hybridized energies $E_{n,-}$ and $E_{n,+}$, defining bonding and anti-bonding solution. Both solutions are degenerate for $\beta\rightarrow\infty$ and $\theta=n\pi$. For increasing $\beta$ and $\theta\neq n\pi$, however, the states $\Psi_{n,+}(x)$ and $\Psi_{n,-}(x)$ separate into different dots, while they exchange their spatial distribution by crossing the points $\theta=n\pi$. This feature will be the key aspect of the chiral anomaly in real space to be specified below.

In order to investigate the charge distribution, we define the charge operator as \cite{Charge}
\begin{equation}
\label{Eq:TotalCharge}
\hat{Q}=\frac{1}{2}\int_{-L/2}^{L/2}dx \left[\hat{\Psi}^{\dagger}(x),\hat{\Psi}(x)\right].
\end{equation}
In addition, we specify the partial charge operator $\hat{Q}_A$ and $\hat{Q}_B$ by
\begin{eqnarray}
\label{Eq:PartialCharge1}
\hat{Q}_{\lambda}=\frac{1}{2}\lambda\int_{0}^{\lambda L/2}dx \left[\hat{\Psi}^{\dagger}(x),\hat{\Psi}(x)\right],
\end{eqnarray}
counting the charge contained in region A, B, respectively.
Eqs. (\ref{Eq:TotalCharge}) and (\ref{Eq:PartialCharge1}) are rewritten in momentum representation, using an eigenfunction expansion in terms of the eigenfunctions of Eq. (\ref{Eq:Ansatz})
\begin{eqnarray}
\label{Eq:PartialCharge3}
\hat{Q}=\sum_{n,\alpha=\pm}\left(\hat{c}_{n,\alpha}^{\dagger}\hat{c}_{n,\alpha}-\frac{1}{2}\right),\\
\label{Eq:PartialCharge4}
\hat{Q}_\lambda=\sum_{n,\alpha=\pm}L\vert a_{\lambda,\alpha}\vert^2\left(\hat{c}^{\dagger}_{n,\alpha}\hat{c}_{n,\alpha}-\frac{1}{2}\right)+\nonumber \\
\sum_{l,j,\alpha=\pm}F_{\lambda,\alpha}(\beta,\theta,j,l)\hat{c}_{j,\alpha}^{\dagger}\hat{c}_{l,-\alpha},
\end{eqnarray}
where $\lbrace n,l,j\rbrace\in\mathcal{Z}$ and the amplitudes of the transfer terms $F_{\lambda,\alpha}(\beta,\theta,l,j)$ are displayed in App. B.

\begin{figure}
\includegraphics[scale=0.4]{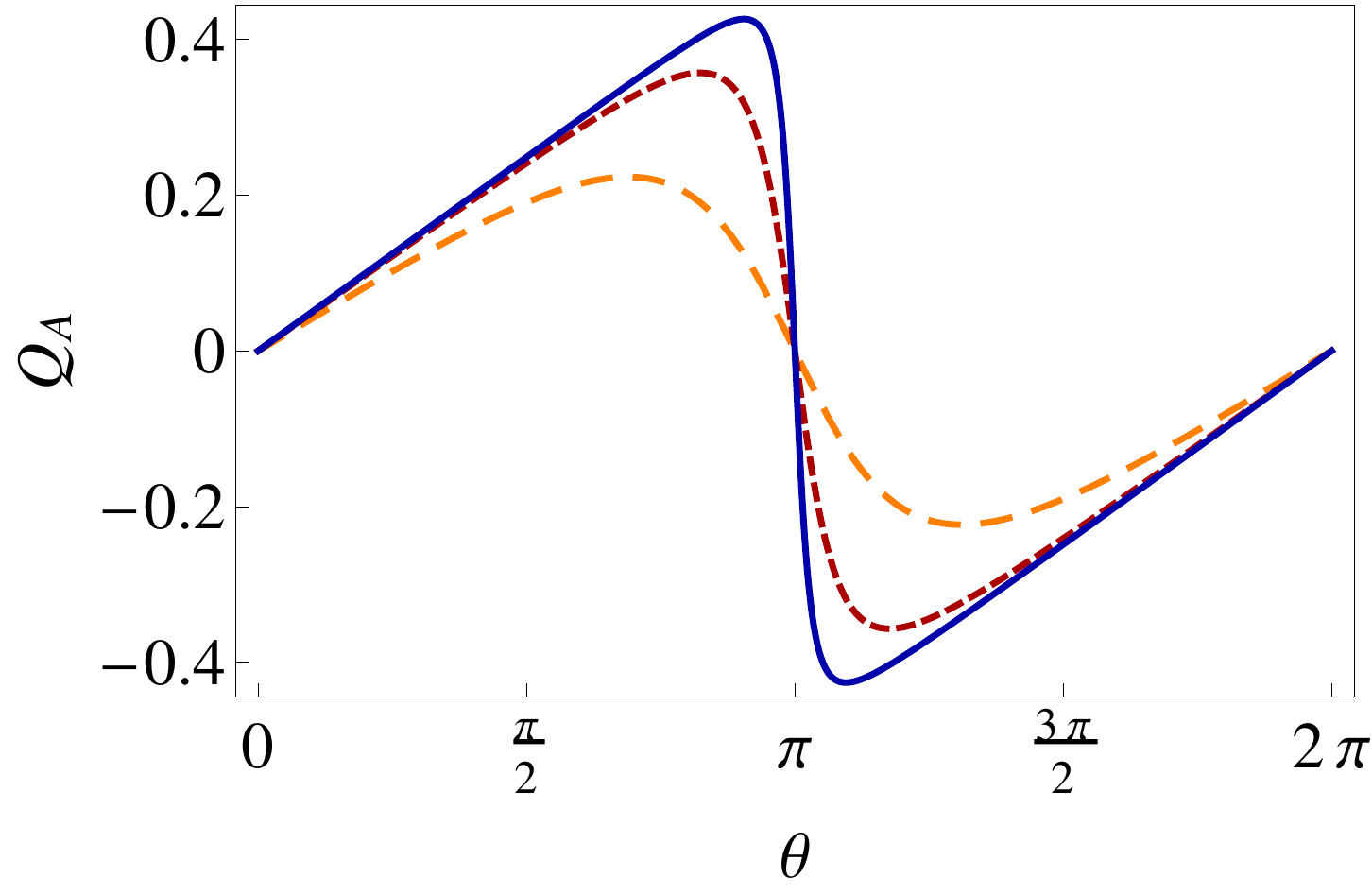}
\caption{(Color online) Average partial charge in region A with $N_+=N_-+1$ in units of the elementary charge $e$ for magnetic strength $\beta=1$ (orange, dashed), $\beta=2$ (red, dashed) and $\beta=3$ (blue, solid) as a function of the magnetization angle $\theta$.}
\label{Fig:PartialCharge}
\end{figure}

To evaluate the expectation values of the total charge and the partial charges of Eqs. (\ref{Eq:PartialCharge3}) and (\ref{Eq:PartialCharge4}) on an arbitrary many particle state, we use the heat-kernel regularization \cite{5, 4}, where every addend of the sum receives a regulator of the form $\exp(-\epsilon \vert E_n\vert /v_F)$
\begin{eqnarray}
\label{Eq:Regularization}
\sum_{n,\alpha=\pm} g_{\alpha}(n)~\rightarrow~\sum_{n,\alpha=\pm} g_{\alpha}(n)e^{-\epsilon \vert E_{n,\alpha}\vert/v_F}.
\end{eqnarray}
Using this replacement and the energy spectrum $E_{n,\alpha}=v_Fq_{n,\alpha}$, we perform the sum of Eqs. (\ref{Eq:PartialCharge3},\ref{Eq:PartialCharge4}) and, eventually, take the limit $\epsilon\rightarrow 0$. The total charge then yields
$\langle \hat{Q}\rangle =N_++N_-$
with $N_\alpha$ being the excess number of particles over the Fermi sea corresponding to the energy dispersion $E_{N,\alpha}$.
The partial charges can likewise be calculated as
\begin{eqnarray}
\label{Eq:PartialCharge8}
\langle\hat{Q}_{\lambda}\rangle=\sum_{\alpha=\pm}L\vert a_{\lambda,\alpha}\vert^2\left(N_{\alpha}+\frac{\phi_{\alpha}(\beta,\theta)}{2\pi}+\frac{1}{2}\right)\; .\nonumber
\end{eqnarray}
The average partial charge, confined in region A, is plotted in Fig. \ref{Fig:PartialCharge}.
We, evidently, observe a $2\pi$-periodicity of the charge, leading to the conservation of the fermion number in a full circle rotation of $\theta$ for any finite $\beta$. In the limit $\beta\rightarrow\infty$, however, we obtain discontinuities at $\theta=(2n+1)\pi$ ( $\theta=2n\pi$, respectively, depending on the filling).

To address the stability of the fractional charges, we evaluate the commutator of the partial charge operators of Eqs. (\ref{Eq:PartialCharge3}) and (\ref{Eq:PartialCharge4}) and the Hamiltonian
\begin{eqnarray}
\label{Eq:PartialCharge6}
\left[\hat{H},\hat{Q}_\lambda\right]=\sum_{l,j,\alpha=\pm}\frac{f_{\lambda,\alpha}(\beta,\theta,l,j)}{i\vert c_+(\beta,\theta)\vert\vert c_-(\beta,\theta)\vert}\hat{c}_{j,\alpha}^{\dagger}\hat{c}_{l,-\alpha}v_F.\nonumber
\end{eqnarray}
where the functions $f_{\lambda,\alpha}(\beta,\theta,l,j)$ and $c_{\alpha}(\beta,\theta)$ are defined in App. B. In the limit $\beta\rightarrow\infty$, the commutator decays to zero exponentially since $f_{\lambda,\alpha}(\beta,\theta,l,j)$ is bounded and $\vert c_{\alpha}(\beta,\theta)\vert\vert c_{-\alpha}(\beta,\theta)\vert\sim\exp(\beta)$. The corresponding fractional charges are, thus, stable quantum observables, interpreted as separated soliton and anti-soliton. In a mapping to the Jackiw-Rebbi model, the magnetic strength $\beta$ would correspond to the spatial distance $L$ between soliton and anti-soliton.
%
%

The description of the eigenstates with the $2\pi$-periodic eigenenergies $E_{n,\alpha}=v_Fq_{n,\alpha}$ implies a spatial distribution in both dots, where the weights of the wave functions in dot A and B exchange by variation of $\theta$ around $n\pi$, provided there is a finite weight in each dot. This situation changes qualitatively for $\beta\rightarrow\infty$. Then, the regions A and B are totally decoupled and particles are confined in each dot separately. The degeneracy of the energy levels $E_{N,+}$ and $E_{N,-}$ for $\theta=\pi$, proposes a new set of eigenenergies for the states located in A ($\alpha=-$) and B ($\alpha=+$), respectively,
\begin{eqnarray}
\label{Eq:EnergyDispersionA}
\tilde{E}_{n,\alpha}=v_F \tilde{q}_{n,\alpha}
\end{eqnarray}
with the quantization of the wavevector $\tilde{q}_{n,\alpha}=(2\pi/ L)\left(\alpha\left(1/2-\theta/(2\pi)\right)+n\right)$, reproducing the results of \cite{1, 2}.
%
The decoupling of the dots lead to twisted boundary conditions also for the inner barrier, implying a coupling of the fields $\hat{\psi}_{\uparrow}$ and $\hat{\psi}_{\downarrow}$. This coupling then allows us to describe each dot by an unfolded Hamiltonian, where we only employ one chiral field $\hat{\psi}_{\alpha}(x)$ in each dot,
\begin{equation}
\label{Eq:HamiltonianDot1}
H_{\alpha}=\int_{-L/2}^{L/2}\alpha\hat{\psi}_{\alpha}^{\dagger}(x)\left(-iv_F\partial_x-\frac{2\pi v_F}{L}\left(\frac{\theta}{2\pi}-\frac{1}{2}\right)\right)\hat{\psi}_{\alpha}(x)
\end{equation}
with
\begin{eqnarray}
\hat{\psi}_{\alpha}(x)=\sum_{n}e^{i \alpha (2\pi/L)nx}\hat{c}_n\nonumber \; .
\end{eqnarray}
The Hamiltonian of the full system is then built by Eq. (\ref{Eq:HamiltonianDot1}) as the sum of the Hamiltonians of the two dots
\begin{equation}
\label{Eq:Mapping2}
H=\sum_{\alpha}H_{\alpha}=\int_{-L/2}^{L/2} dx \hat{\mathcal{X}}^{\dagger}(x)\left(-i\partial_x-eA\right)v_F\sigma_z\hat{\mathcal{X}}(x)
\end{equation}
with $\hat{\mathcal{X}}(x)=\left(\hat{\psi}_+(x),\hat{\psi}_-(x)\right)^T$ and the effective electro-magnetic vector potential $A=(2\pi/eL)(\theta/(2\pi)-1/2)$. Eq. (\ref{Eq:Mapping2}) represents the Hamiltonian of a QSH ring with periodic boundary conditions under the influence of an external electro-magnetic field \cite{6}.

\begin{figure}
\centering
\includegraphics[scale=0.45]{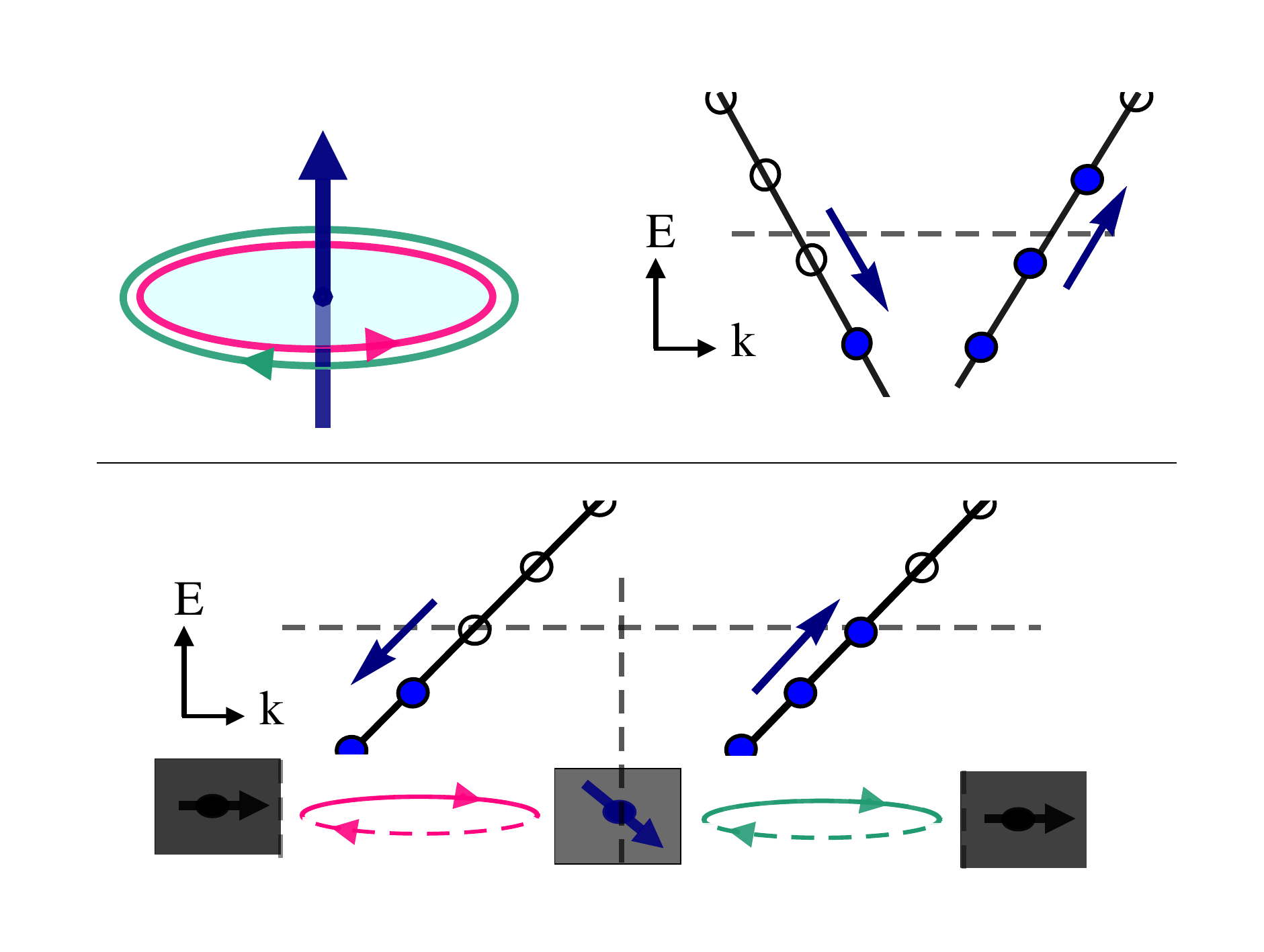}
\caption{(Color online) Schematic representation of the system, mapped onto a QSH ring (upper half) and the initial impurity system (lower half). A filled ball indicates an occupied state, while circles denote unoccupied states. The arrows beside the dispersion denote the evolution in a time dependent variation of $\theta$.}
\label{Fig:ChiralAnomalyRealSpace}
\end{figure}

Following Ref. \cite{6}, we are able to treat Eq. (\ref{Eq:Mapping2}) in linear response for $\theta\sim\pi$ to compute the average of the charge contained in one dot for that configuration
\begin{equation}
\label{Eq:LinearResponse}
\langle\hat{Q}_{\alpha}\rangle=\tilde{N}_{\alpha}-\alpha
\frac{\theta}{2\pi}.
\end{equation}
Induced by the effective electro-magnetic vector potential, we find additional contributions to the background charge $\tilde{N}_{\alpha}$, counting the vacuum fluctuations of the underlying Fermi sea. Eq. (\ref{Eq:LinearResponse}) implies that we can change the number of particles in each dot by a variation of $\theta$. Thus, due to spectral flow, chiral currents are not conserved. Indeed a quantum field theoretical description in 1+1 dimension shows \cite{KFuji, RBertl, MPeskin} that
\begin{eqnarray}
\label{Eq:ChiralCurrent}
\partial_{\mu}J^{\mu}_5 = - \frac{e}{2\pi}\epsilon^{\mu\nu}F_{\mu\nu},\nonumber
\end{eqnarray}
where $F_{\mu\nu}$ represents the electro-magnetic field tensor, $\epsilon^{\mu\nu}$ is the total antisymmetric tensor, $e$ the elementary charge and $J^{\mu}_5$ is the chiral current in 1+1 dimension obtained by $J^0_5=-\hat{\mathcal{X}}^{\dagger}\sigma_z\hat{\mathcal{X}}$,
and $J^{1}_5=-\hat{\mathcal{X}}^{\dagger}\hat{\mathcal{X}}$.
For our case, we find $A_0=0,~~A_1=(2\pi/eL)(\theta/(2\pi)-1/2)$, leading to $\partial_{\mu}J^{\mu}_5=-\partial_t \theta(t)/(\pi v_F L)$. The system of the three magnetic impurities with sufficiently large magnetic barrier strength thus witnesses the chiral anomaly in real space for a time-dependent variation of $\theta$. Fig. \ref{Fig:ChiralAnomalyRealSpace} schematically illustrates the connection to real space. On the mapped system of the QSH ring, the chiral anomaly is visible in momentum space, while in the original system, the branches are separated between the dots. Meaning, particle numbers in dot A and B are not conserved in a time dependent variation of $\theta$, even though they are confined to each dot, separately. The chiral anomaly appears as a peculiarity of the one dimensional Dirac equation, where the quantum fluctuations of right and left movers are coupled. However, our system is neither unbounded nor purely one dimensional. The one dimensional edge states in each dot eventually merge into two dimensional bulk states connecting the dots and implementing the physical mechanism of broken chiral symmetry.

\begin{figure}
\includegraphics[scale=0.5]{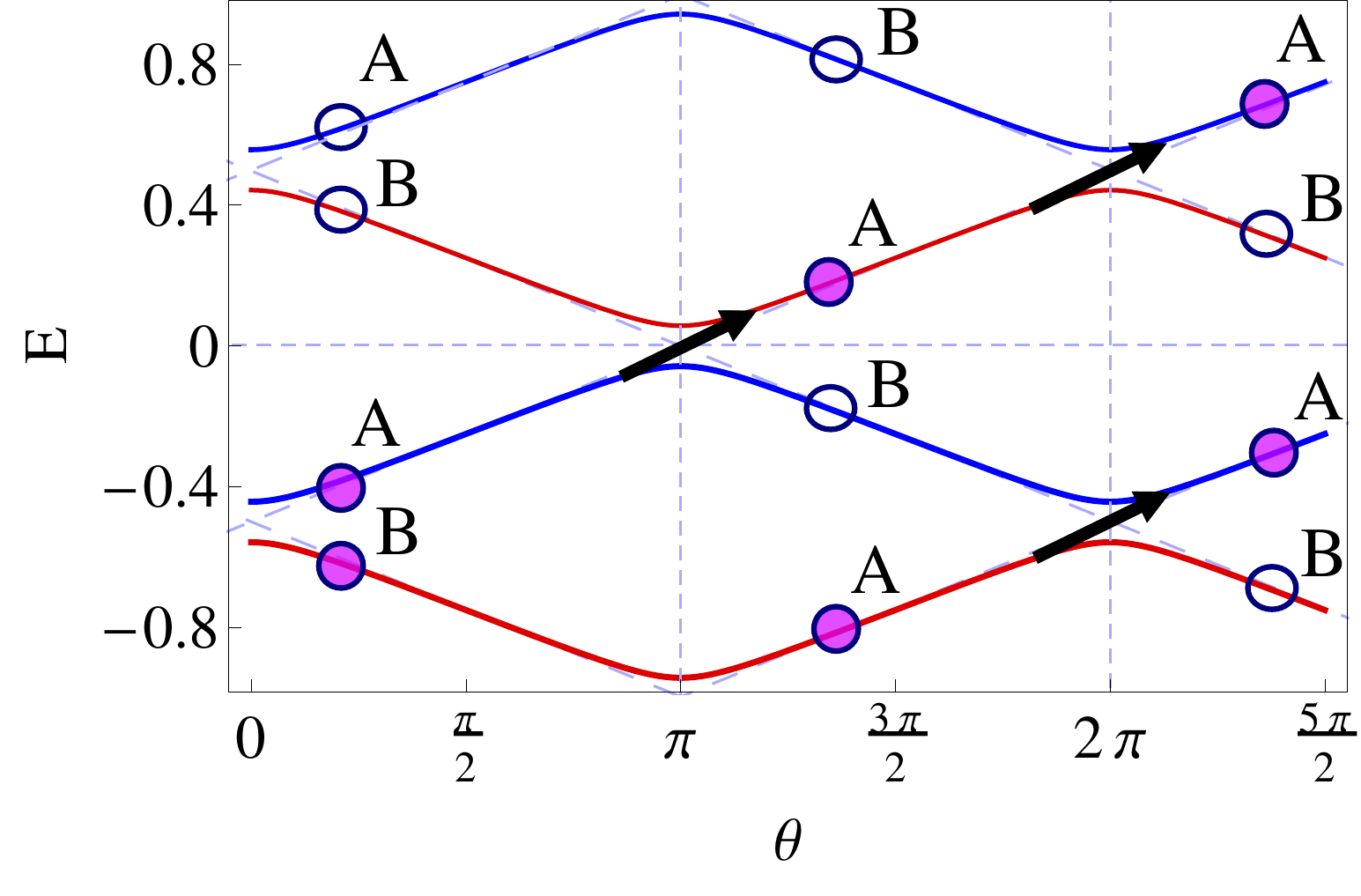}
\caption{(Color online) Energy levels $E_{0,-}$ ,$E_{1,-}$ (solid, blue) and $E_{-1,+}$, $E_{0,+}$ (solid, red) pictured for $\beta=1.7$ in units $2\pi v_F/L$. A filled circle denotes a occupied state, while an unfilled tags an unoccupied state. The index A, B, respectivley, besides the circles, corresponds to the region of main spatial distribution of the corresponding state.}
\label{Fig:LandauZener}
\end{figure}

Evidently, the chiral anomaly in real space is present in our system for large (infinite) magnetization of the impurities. In the case of a smaller (finite) magnetization, however, we obtain a $2\pi$ periodicity of the occupation of the dots as a function of $\theta$. Thus, for an adiabatical, i.e. slow, manipulation of $\theta$, particle numbers are conserved in full circle rotations, and the chiral anomaly is absent. At first sight, this looks like a dilemma for our proposal, but, fortunately, we have found a way out: For a large (but finite) value of $\beta$, the weight of the wave functions is exponentially small in one dot. Hence, a non-adiabatical manipulation of $\theta$, allows us to preserve the spatial distribution. The reason is that we can overcome the anti-crossings of the energy levels if $\theta$ is varied fast enough, {\it{cf.}} Fig. \ref{Fig:LandauZener}.
The physics of this transition is the one of the Landau-Zener problem \cite{7, 8, 9}.
Assuming a linear time dependence of $\theta=\gamma t$, the transition probability to overcome the emerging energy gap at $\theta=n\pi$ is then given by
\begin{equation}
\label{Eq:Probability}
P=\exp\left(-C(\beta)\right).
\end{equation}
with $C(\beta)=\pi v_F/\left[L\gamma\sinh^2(\beta)\right]$.
In the unbounded spectrum, an approach of energy levels appears with a periodicity of $\pi$ for all energy levels (see Fig. \ref{Fig:LandauZener}). The probability of an excitation, however, only contributes in an exchange process of an occupied with an unoccupied state at the Fermi surface. We find a closed expression for the probability that all states remain in their spatial configuration, by performing a rotation of $\Delta \theta=N2\pi$
\begin{eqnarray} \label{Eq:LandauZenerProbability4}
P(N)&=&\prod_i^{2N}P^i=\exp\left(-C(\beta)\right)^{N+2N^2}
\end{eqnarray}
After a rotation of $\Delta\theta=N2\pi$, we have populated states in dot A and depopulated states in dot B (see Fig. \ref{Fig:LandauZener}).
Thus, with a probability of $P(N)$ we change the particle number in region A and B without directly shifting particles from one dot to the other one which comes from the chiral anomaly in real space.

In conlusion, we have demonstrated the emergence of stable fractional charges, interpreted as soliton and anti-soliton states, at the helical edge in the presence of three magnetic impurities. By mapping this system onto a helical ring, pierced by a magnetic flux, we have demonstrated that the chiral symmetry is broken for a time-dependent variation of the relative orientation $\theta$ of the magnetization of the center impurity. This setup enables us to introduce the chiral anomaly into real space. If the change in the number of charges in each quantum dot formed at the helical edge could be monitored, e.g. by a proximate single electron transistor, then this signal -- in the presence of a time-dependent variation of $\theta$ -- would be a direct evidence of the chiral anomaly.

\begin{acknowledgments} Financial support by the DFG (SPP1666 and SFB1170 "ToCoTronics"), the Helmholtz Foundation (VITI), and the ENB Graduate school on "Topological Insulators" is gratefully acknowledged. We thank F.\ Cr\'epin for interesting discussions.
\end{acknowledgments}

\newpage

\section{Appendix}
\subsection{Appendix A}
The investigated system of the three magnetic impurities yields boundary conditions at the points $x=0,\pm L/2$. Applying the methods of Refs. \cite{1, 2}, we obtain
\begin{eqnarray}
\label{Eq:BC1}
\left(1\pm\sigma_y\right)\Psi(\mp L/2)=0.
\end{eqnarray}
Moreover, we identify a boundary condition at $x=0$, conneting the regions A and B in the case of finite magnetic strength $m$, by
\begin{equation}
\label{Eq:BC3}
\begin{split}
\Psi(0^+)=\lbrace\cosh(m/v_F)\\
-\sinh(m/v_F)\left(\sigma_y
\cos(\theta)+\sigma_x\sin(\theta)\right)\rbrace\Psi(0^-).
\end{split}
\end{equation}
Using Eqs. (\ref{Eq:BC1}) and (\ref{Eq:BC3}) as well as the ansatz
\begin{eqnarray}
\label{AppendixEq:Ansatz}
\Psi_{n,\alpha}(x)=\Psi_{n,\alpha,A}(x)\theta(-x)\theta(x+L/2)\nonumber\\
+\Psi_{n,\alpha,B}(x)\theta(x)\theta(-x-L/2)
\end{eqnarray}
with
\begin{eqnarray}
\label{AppendixEq:Ansatz1}
\Psi_{n,\alpha,\lambda}(x)=\begin{pmatrix}
a_{\lambda,\alpha}e^{iqx}\\
b_{\lambda,\alpha}e^{-iqx}\\
\end{pmatrix},
\end{eqnarray}
we have one free parameter in calculating the coefficients $a_{\lambda,\alpha}$ and  $b_{\lambda,\alpha}$. With no loss of generality, we choose $a_{A,\alpha}=1/c_{\alpha}(\beta,\theta)$, with the normalization factor $c_{\alpha}(\beta,\theta)$ and $\alpha=\pm$. Straight forward algebra leads to the solutions
\begin{eqnarray}
\label{Eq:Parameterb}
b_{A,\alpha}&=&\frac{-i\exp(-i\phi_{\alpha}(\beta,\theta))}{c_{\alpha}(\beta,\theta)},\nonumber\\
a_{B,\alpha}&=&\frac{\cosh(\beta)+\sinh(\beta)\exp\lbrace i\left(\theta-\phi_{\alpha}(\beta,\theta)\right)\rbrace}{c_{\alpha}(\beta,\theta)},\nonumber\\
b_{B,\alpha}&=&\frac{-i\left\lbrace\sinh(\beta)\exp\left(-i\theta\right)+\cosh(\beta)\exp\left(-i\phi_{\alpha}(\beta,\theta)\right)\right\rbrace}{c_{\alpha}(\beta,\theta)}.\nonumber
\end{eqnarray}
The quantity $\phi_{\alpha}(\beta,\theta)$ is part of the quantization of $q_{n,\alpha}$ (see Eq. (\ref{Eq:Quantization3Impurities}) of the main text).
The normalization constants $c_{\alpha}(\beta,\theta)$ can be calculated as
\begin{eqnarray}
c_{\alpha}(\beta,\theta)=\sqrt{(1+h_{\alpha}(\beta,\theta))L}\nonumber
\end{eqnarray}
with $h_{\alpha}(\beta,\theta)$ given by
\begin{eqnarray}
h_{\alpha}(\beta,\theta)=\frac{1}{2}\lbrace\cosh^2(\beta)+\sinh^2(\beta)+\nonumber\\
\sinh(\beta)\cosh(\beta
)\cos(\theta-\phi_{\alpha}(\beta,\theta))\rbrace.\nonumber
\end{eqnarray}
\subsection{Appendix B}
To calculate the commutator of the partial charge operators
\begin{eqnarray}
\label{AppendixB1}
\hat{Q}_{\lambda}=\frac{1}{2}\lambda\int_{0}^{\lambda L/2}dx \left[\hat{\Psi}^{\dagger}(x),\hat{\Psi}(x)\right],
\end{eqnarray}
and the Hamiltonian, we first rewrite Eq. (\ref{AppendixB1}) in momentum representation, using an eigenfunction expansion of the field operators in terms of the wave functions specified in App. A
\begin{equation}
\label{AppendixB2}
\hat{Q}_{\lambda}=\lambda\int_{0}^{{\lambda}L/2}dx~\frac{1}{2}\sum_{n,m,\alpha,\beta}\Psi_{n,\alpha}^{\dagger}(x)\Psi_{m,\beta}(x)\left[\hat{c}^{\dagger}_{n,\alpha},\hat{c}_{m,\beta}\right].
\end{equation}
This directly leads to the form
\begin{equation}
\begin{split}
\hat{Q}_{\lambda}=\sum_{n,\alpha}L\vert a_{\alpha}\vert^2\left(\hat{c}^{\dagger}_{n,\alpha}\hat{c}_{n,\alpha}-\frac{1}{2}\right)+\\
\sum_{n,m,\alpha}\lambda\int_{0}^{\lambda L/2}dx~\Psi^{\dagger}_{n,\alpha}(x)\Psi_{m,-\alpha}(x)\hat{c}^{\dagger}_{n,\alpha}\hat{c}_{m,-\alpha}.
\end{split}
\end{equation}
We define now
\begin{equation}
F_{\lambda,\alpha}(\beta,\theta,n,m)=\lambda\int_{0}^{\lambda L/2}dx~\Psi^{\dagger}_{n,\alpha}(x)\Psi_{m,-\alpha}(x).
\end{equation}
With the use of the explicit eigenfunctions $\Psi_{n,\alpha}(x)$, we obtain the functions $F_{\lambda,\alpha}(\beta,\theta,l,j)$ to be
\begin{eqnarray}
F_{\lambda,\alpha}(\beta,\theta,j,l)&=&\frac{L}{2\phi_{\alpha}-2\pi(l-j)}\frac{f_{\lambda,\alpha}(\beta,\theta,j,l)}{i\vert c_+(\beta,\theta)\vert\vert c_-(\beta,\theta)\vert}\nonumber\\
\end{eqnarray}
with $l,j\in\mathcal{Z}$. The function $f_{\lambda,\alpha}(\beta,\theta,l,j)$ are given by
\begin{eqnarray}
f_{A,\alpha}(\beta,\theta,j,l)&=&e^{2i\phi_{\alpha}}-1 \; , \nonumber\\
f_{B,\alpha}(\beta,\theta,j,l)&=&2ie^{i\phi_{\alpha}}\bigg[2
\sin(\phi_{\alpha})\sin^2\left(\frac{1}{2}(l-j)\pi+\frac{1}{2}\phi_{\alpha}\right) \nonumber\\
&+&\sin\left((l-j)\pi+\phi_{\alpha}\right)\big(\cos(\theta)\sinh(2\beta)
\nonumber\\
&+&\cos(\phi_{\alpha})\cosh(2\beta)\big)\bigg] \; .
\end{eqnarray}
To evaluate now the commutator of the Hamiltonian and the partial charge operators of Eq. (\ref{AppendixB2}), we first write the Hamiltonian in momentum representation, i.e.
\begin{equation}
\hat{H}=\sum_{n,\gamma=\pm}p_{n,\gamma}\hat{c}_{n,\gamma}^{\dagger}\hat{c}_{n,\gamma}.
\end{equation}
Thus, the commutator of $\hat{H}$ and the partial charge operators of Eq. (\ref{AppendixB2}) read
\begin{equation}
\label{Eq:PartialCharge3MagImp5}
\begin{split}
\left[\hat{H},\hat{Q}_{\lambda}\right]=\sum_{n,m,\alpha\neq\beta}\sum_{l,\gamma}\lambda\int_{0}^{\lambda L/2}dx~\Psi^{\dagger}_{n,\alpha}(x)\Psi_{m,\beta}(x)v_F\\
\left[p_{l,\gamma}\hat{c}_{l,\gamma}^{\dagger}\hat{c}_{l,\gamma},\hat{c}^{\dagger}_{n,\alpha}\hat{c}_{m,\beta}\right].
\end{split}
\end{equation}
The remaining commutator of fermionic creation and annihilation operators yields
\begin{equation}
\label{Eq:PartialCharge3MagImp6}
\begin{split}
\left[p_{l,\gamma}\hat{c}_{l,\gamma}^{\dagger}\hat{c}_{l,\gamma},\hat{c}^{\dagger}_{n,\alpha}\hat{c}_{m,\beta}\right]=\\
-p_{l,\gamma}\left(\hat{c}_{n,\alpha}^{\dagger}\hat{c}_{l,\gamma}\delta_{l,m}\delta_{\beta,\gamma}-\hat{c}_{l,\gamma}^{\dagger}\hat{c}_{m,\beta}\delta_{l,n}\delta_{\alpha,\gamma}\right).
\end{split}
\end{equation}
Inserting Eq. (\ref{Eq:PartialCharge3MagImp6}) into Eq. (\ref{Eq:PartialCharge3MagImp5}), we obtain
\begin{eqnarray}
\left[\hat{H},\hat{Q}_{\lambda}\right]=\sum_{n,m,\alpha\neq\beta}\lambda v_F\int_{0}^{\lambda L/2}dx~\Psi^{\dagger}_{n,\alpha}(x)\Psi_{m,\beta}(x)\nonumber\\
\hat{c}_{n,\alpha}^{\dagger}\hat{c}_{m,\beta}(p_{n,\alpha}-p_{m,\beta}).\nonumber
\end{eqnarray}
Furthermore, we use that
\begin{eqnarray}
\left(p_{n,\alpha}-p_{m,\beta}\right)\frac{L}{2}=\phi_{\alpha}-\pi(n-m),\nonumber
\end{eqnarray}
and finally we reach at
\begin{eqnarray}
&& \left[\hat{H},\hat{Q}_\lambda\right] \nonumber\\
&& =\sum_{n,m,\alpha=\pm}F_{\lambda,\alpha}(\beta,\theta,n,m)(2\phi_{\alpha}-2\pi(n-m))/L \nonumber\\
&& =\sum_{n,m,\alpha=\pm}\frac{v_F f_{\lambda,\alpha}(\beta,\theta,n,m)}{i\vert c_+(\beta,\theta)\vert\vert c_-(\beta,\theta)\vert}\hat{c}_{n,\alpha}^{\dagger}\hat{c}_{m,-\alpha}.\nonumber
\end{eqnarray}
The commutators only vanish, if all elements of the sum vanish, i. e.
\begin{eqnarray}
\label{Eq:PartialCharge7}
f_{\lambda,\alpha}(\beta,\theta,n,m)=0.\nonumber
\end{eqnarray}
With the explicit form of $f_{\lambda,\alpha}(\beta,\theta,n,m)$, for all $n,m\in\mathcal{Z}$, we then obtain the solutions
\begin{eqnarray}
\theta=\pm\arccos\left(\pm\coth(\beta)\right).\nonumber
\end{eqnarray}
For any finite $\beta$, the $\coth$ yields values larger than unity. Thus, the arccos yields a complex number. In the limit $\beta\rightarrow\infty$, the coth becomes unity leading to a real angle. In that special case, however, the commutator vanishes for a different reason caused by the divergence of $\vert c_{\alpha}(\beta,\theta)\vert \vert c_{-\alpha}(\beta,\theta)\vert\sim\exp(\beta)$.

\begin{thebibliography}{99}
\bibitem{beta} Note that $\beta$ is dimensionless. It quantifies the magnetude of the energy scale of the center impurity $m/L$ with respect to the Thouless energy $v_F/L$.
\bibitem{Charge} This definition of the charge is commonly used in particle physics because of the pecularities of the unbounded spectrum.
\end{thebibliography}
\end{document}